\documentclass[showpacs, showkeys ,preprintnumbers]{revtex4}
\usepackage{graphicx}
\usepackage{dcolumn}
\usepackage{bm}

\begin{document}
\title{Fairness State with Plastic Preferences}
\author{Elena Ramirez Barrios}
\author{Juan G. Diaz Ochoa}
\email{diazochoa@itp.uni-bremen.de} 
\affiliation{\mbox{Department of Economics, University of Kiel,
  Postal Address 1 (Wilhelm Seelig Platz 1, D-24098 Kiel, Germany)}}
\affiliation{\mbox{Institute for physics,Fachbereich 1, Bremen University,
Otto Hahn Alle, D28334 Bremen, Germany}}
\begin{abstract}
The definition of preferences assigned to individuals is a concept that concerns many disciplines, from economics, with the search of an acceptable outcome for an ensemble of individuals, to decision making an analysis of vote systems. We are concerned in the phenomena of good selection and economic fairness. In Arrow's theorem this situation is expressed as the impossibility of aggregate preferences among individuals falling down into some unfairness state. This situation was also analyzed in a previous model in a network of individuals with a random allocation \cite{Ramirez}. Both analysis are based on static preferences. 

In a real society the individuals are confronted to information exchange which can modify the way that they think. Also, the preference formation of each individual is influenced by this exchange. This consideration reveals why the actual theory is not able to make an accurate analysis of the influence of the individual, or cluster of individuals, in the fairness state. The aim of this investigation is to consider the coupling of two systems, one for the formation of preferences and a second where an allocation of goods is done in an evolutionary environment. 
\end{abstract}
\pacs{89.65.Gh; 89.75.-k; 89.65.-s}
\keywords{Goods allocation; Fairness; Envy; Agent based models}
\maketitle
\section{Introduction}
Commonly, envy is considered as a feeling with a negative character that affects the social relation between individuals. The measure of the influence of this feeling into the social systems, i.e., the way that it influence the choice and the change of preferences in time, can be explained from a behavioristic point of view explaining, by means of different approaches, the influence of feelings at the moment of choice. Emotions act as a motor for consume and social relations help to improve the satisfaction level and adjust processes can be considered as important ingredients to achieve low envy levels in the society.

In conventional models the preferences are constant in space and time. The individuals are able to set an own preference list with a particular rank in order to achieve the highest satisfaction level. Using these lists they orient themselves in order to get a bundle of goods. We avoid the problem of prices by introducing an assignation mechanism. When an individual does not acquire her/his desired goods, then she/he start to looks into the neighbor's basket. When the individual find out the searched good into her/his neighbor's basket an envy relation is bearing \cite{Ramirez}. The number of comparisons is unlimited and is used as a measure of time. Therefore, we can state that envy relations have a dynamical nature.

This dynamic takes place in a \textit{behavioral sphere}, just before any transaction between individuals, and its presence affects, beyond the market, the existent economical and social systems. The emergence of envy, among other feelings, between individuals provide an ideal field which allows the exchange of goods \cite{Sartorius}\cite{Rawls}. Following Simon \cite{Simon}, the individuals are doted with bounded rationality and doesn't believe in preference or utility relations. That means, the choice is not deterministic and normally, incomplete information is present. The develop of own beliefs to improve the choice has a temporal sense. Furthermore, talk about preference or utility has no sense, and the idea is better related with satisfaction. The change of believes has no dependence on typical scales of time, but on times of interpersonal comparisons.

In the real world, preferences and satisfaction levels change in time due externalities (i.e., Marketing, fashion, introduction of new products, but also social interaction as cooperation, and other issues as altruism or donations) and internal or behavioral influences -Menger's proposition as explanation of changing and evolution of preferences (Menger, 1950)-, demand for innovations\cite{Witt_I}, demonstrated dissatisfaction with consumed goods, desire of differentiation, and possession of a veil of ignorance \cite{Rawls}, individuals must permanently change the way they choose. Therefore, a kind of 'elastic preferences' must be introduced in order to achieve the highest satisfaction level, allowing its change in time.

In this investigation we show a slight modification of a previous model \cite{Ramirez} and propose a system where the individuals try to adapt their preferences in order to avoid envy, like a system that adapts itself to the environment and will avoid posterior exchange of goods. The second part provides the model and consider plastic preferences into the baseline of the envy mechanism. Additionally, we explain the simulation method we developed. In the third part we discuss our results. Main conclusions and outlook are provided in the last part.
\section{Model}
\subsection{Rigid preferences}
Principally, we need three elements to built our model: the first one is a set of various goods of $K$ different kinds located in a 'store'. For each kind $k$ of goods, a specific number $G(k)$ of goods exists. Our second element is a set of individuals. There are $N$ individuals in our model. Each individual $i$ has a preference list $P_i$ of her/his preferred goods. This preference list $P_i$ can be coded as a permutation of the numbers $1,\dots,K$ with $P_i(1)$ being most important good for individual $i$, $P_i(2)$ being the second important good, and so on. The third element is a set of individual 'baskets' $B_i$ in which the individuals can store their goods after picking them up from the set of goods in the store. Thus, $B_i(k)$ denotes the number of goods of kind $k$ the individual $i$ has got in her/his basket.

According to their individual preferences, the individuals search for goods of their highest ranked preference type in the available set of goods. Every individual is allowed to take an overall number $M$ of goods. If the highest preferred good is no more available, the individual starts to collect the second good on her/his preference list, and so on. Each individual takes the goods from the set of goods into her/his basket. Several individuals enter the store in a random order. The first individual according to this randomly created queue takes $M$ stored goods (for example motorbikes, computers, etc.) into her/his basket. Then she/he leaves the store and the second individual in the queue enters the store. She/he also takes $M$ goods, selecting them according to the preference list she/he has in mind, and leaves the store. This is repeated until the $N$ individual searches into the remaining $M$ goods. After all the goods have been picked up by the individuals, the following equation holds 
\begin{equation}
\sum_{k=1}^K G(k) = \sum_{i=1}^N \sum_{k=1}^K B_i(k) + R
\end{equation}
with $G(k)$ being the number of goods of kind $k$ in the store before the first individual has entered the store to take some goods into her/his basket. $R$ are the remaining goods in the store that any body wants. The existent number of goods is just sufficient for the individuals in our model and can also be written as
\begin{equation}
N \times M + R= \sum_{k=1}^K G(k) .
\end{equation}
There is no production or destruction of goods. The number of individuals also remains constant. So, when the individual takes a good into her/his basket, there is a depletion of one good in the store.

We define the distribution of the number of goods $G(k)$ according to a distribution function that is more or less uniform. The best suited function in this case is a Gamma function \cite{Ramirez}\cite{Greene} defined by
\begin{equation}
G(k)[A] = \frac{(x-k_{0})^{A-1}e^{-(x-k_{0})}}{\Gamma(A)},
\end{equation} 
where  $k_{0}$ is the index of the good that is peaked around the distribution and $\Gamma(A)$ is the gamma function. In this equation the parameter $A$ is the amplitude of the Gamma distribution, i.e. the variation of this parameter produces a more or less uniform initial distribution of goods in the store. 

After the distribution of goods, the individuals are allowed to take a look into the baskets of some other individuals and to compare their own goods with the goods of these other individuals according to their own preference lists. Some individuals are satisfied because the contain of their baskets fit their own preference lists, while other individuals are missing goods to reach this satisfaction. In this case, envy emerges. Formally, envy can be defined in the following way: two individuals $i$ and $j$ are randomly chosen. They are allowed to look into the baskets of each other. Of course, they first check for the goods on top of their preference lists. Now if individual $i$ sees that $B_j(P_i(1))>B_i(P_i(1))$, i.e., that individual $j$ has more of the good on top of her/his preference list than himself/herself, then individual $i$ feels an emergence of envy. Contrarily, if $B_j(P_i(1))<B_i(P_i(1))$, i.e., if individual $j$ has less of the good on top of the preference list of individual $i$, then individual $i$ is satisfied. In the case that $B_j(P_i(1))=B_i(P_i(1))$, individual $i$ checks for the second good on the preference list. Here again individual $i$ might get satisfied if $B_j(P_i(2))<B_i(P_i(2))$, envious if $B_j(P_i(2))>B_j(P_i(2))$, or willing to inspect the basket according to good $P_i(3)$ if $B_j(P_i(2))=B_i(P_i(2))$. This process is repeated till the baskets are equally filled with goods of higher ranking. In the special case that both baskets are identical, no envy occurs. We can write this envy relation of the individual $i$ might feel towards individual $j$ formally as
\begin{equation} \label{Formel}
E_i(j) = \sum_{k=1}^K \Theta(b_j(P_i(k))-b_i(P_i(k))) \times
\prod_{l=1}^{k-1} \delta(b_i(P_i(l)),b_j(P_i(l)))
\end{equation}
with $\Theta(x)$ the Heaviside function and $\delta(x,y)$ the Kronecker symbol. Note that the addend for $k=1$ in Eq.\ (\ref{Formel}) is well-defined as the empty product $\prod_{l=1}^0 \dots$ has a value of 1.

At the same time when individual $i$ looks into the basket of individual $j$, individual $j$ looks into the basket of individual $i$. Given that the initial preference lists $P_i$ and $P_j$ of individuals $i$ and $j$ are different, three different cases can occur: either none of them feels envy, as both of them are satisfied due to their different preference lists, or exactly one of them feels a local emergence of envy, while the other one feels satisfied, or both are envious towards each other.

Each individual is allowed to make $f$ ``visual contacts'' with other individuals  looking into their baskets. The random selection of pairs of individuals leads to a random network \cite{Bela}. Please note that all edges in this network are undirected, i.e., if there is an edge from individual $i$ to individual $j$, also the edge from $j$ to $i$ exists. This random network can simply be described via a symmetric edge matrix $\eta$, where $\eta(i,j) = 1$ if an edge between $i$ and $j$ exists an 0 otherwise. 

After the edges have been chosen, we can define a Hamiltonian $E$ for this network of individuals, summing up the amount of envy occurring in this network:
\begin{equation}
E=\sum_{i=1}^N \sum_{j=1}^N \eta(i,j) \times E_i(j)
\end{equation}
Note that the value of $E$ is usually different from the number $N_E$ of envious individuals, which is given by
\begin{equation}
N_E=\sum_{i=1}^N \Theta\left(\sum_{j=1}^N \eta(i,j) \times E_i(j)\right) ,
\end{equation}
as a 'dog in the manger' can feel envy towards more than only one other individual.
An envy network is created due to the assignment of goods to the baskets of the individuals, who make comparisons based on their own preferences, and emerges only when the system is out of a Walrasian equilibrium, i.e. when there is a bad allocation of goods. Our model is not a typical optimization problem, in which the global optimum of a proposed pay-off function (free energy for instance) has to be found. When individuals try to find a solution which is optimal for themselves, this solution, which is called Nash equilibrium, might not be a global optimum of the whole problem \cite{Johannes}. Here, our pay-off function, which determines the amount of envy, is a measure of how far the system is from the equilibrium state. So, we analyze the connectivity dependence on the assignation of goods and the number of individuals that express envy. 

Search the fairness state, and not the topology of the network, is the main problem in this investigation \cite{Pastor}. Our particular interest is the measurement of the number $N_E$ of individuals with envy. For this reason, envy is rescaled by the total number of edges in the network divided by the total number of nodes (individuals). There is a fairness state when there are no connections between the nodes of the random network, then $E =0$. Otherwise, we recognize the emergence of an envy state. This network represents the bidirectional exchange of information between individuals, i.e., each individual looks into the basket of the other individual searching for goods of her/his own preference. Because they can look at but not remove goods from the basket of the other individual or exchange goods, the envy relation due to interpersonal comparisons cannot be resolved.

\subsection{Plastic preferences}

Now we introduce the possibility that individuals adjust permanently their preference lists after compare the obtained bundle by a first distribution, when the desires satisfaction level is not achieved. Two fundamental variants can be adopted: either the individuals set complete new preferences, or learn and try to adapt themselves their behavior. The first one corresponds to a random way to change the way to think, whereas the second imply a learn process that imply the possibility to recover information of previous allocation processes. In both cases, the formation of new preferences evolves, conform the system also evolves and does not imply the exchange of information between individuals, or between the store and the individuals, before the allocation process takes place. We studied the first variant because we want to understand the effect that an impulsive individual has into the envy network.

This process implies the introduction of an additional transition probability for the preference list into the whole dynamics of the system. Therefore, a second preference list is randomly created after the individual compare the obtained bundle with her/his neighbors in order to improve her/his obtained bundle. If her/his neighbor has the good that the individual wants to have, but she/he could not obtain, the individual sets a new preference list. The process of setting preferences lists can be done only once before the new distribution, when individuals are able to check their own satisfaction level and compare the obtained bundles in between one more time. A new envy relation can be established if the individuals are not satisfied and must go to search again the source that can remove the deprivation feeling. 

\begin{figure}[th]
\begin{center}
\includegraphics[clip, width=0.6\textwidth]{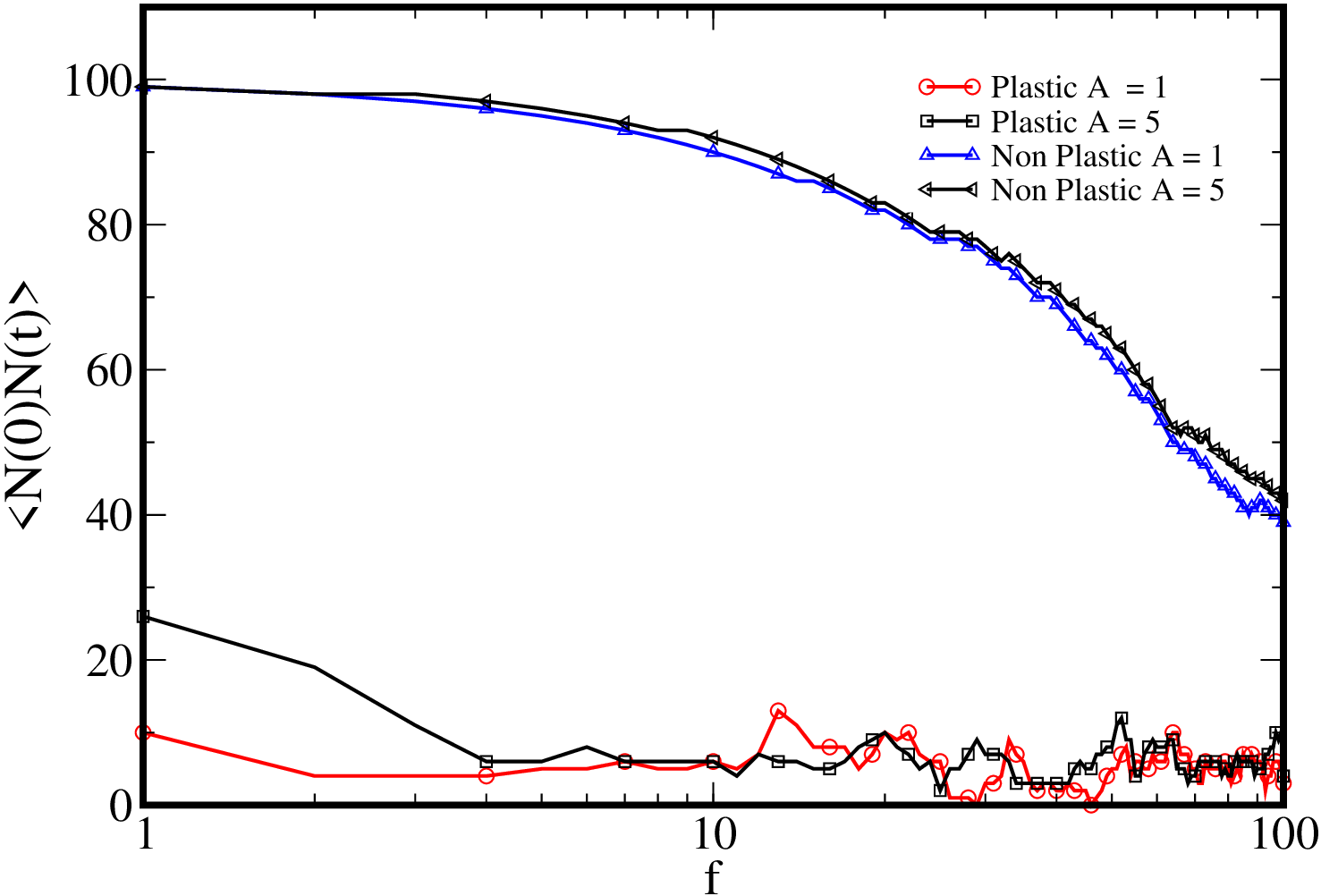}
\end{center}
\caption{Correlation of the number of individuals that express envy as a function of the comparisons times $f$, for individuals with plastic preferences (non normalized). We compare this result with a typical relaxation process for individuals with non plastic (rigid) preferences.}
\label{N2}
\end{figure}

The initial conditions assumes that all the individuals are in a non-envy-state. When the assignation starts, and the preferences are non plastic, then a relaxation process in the correlation function of the number of envious individuals takes place through envy interactions and the number of envious individuals looses the correlation with the initial state. For infinite times the system eventually reaches an equilibrium state. In a social system this process can be seen as a group of individuals meeting their selves with empty baskets and lists of preferences in a market. After a distribution process, the individual test their satisfaction level comparing the acquired goods with their preference lists. In order to avoid envy relations, the switch-on of the plastic preferences introduces  an additional interaction represented by a transition of the preferences of the individuals. The effect of this transition process is observed as a kind of additional force that introduces an arrest into the system and an over-relaxation in the correlation function of envious individuals (See Fig. \ref{N2}).

\section{Results}

We performed simulations with $N=100$ individuals and $K=100$ different kinds of goods. Given that the system is closed (there are no changes in the numbers of goods and the preferences remain the same) then the system must reach an equilibrium state in the distribution of envy and envious individuals for $f\rightarrow \infty$, i.e. we assume a system where envy can 'diffuses' in a set of individuals.

\begin{figure}[th]
\begin{center}
\includegraphics[clip, width=0.6\textwidth]{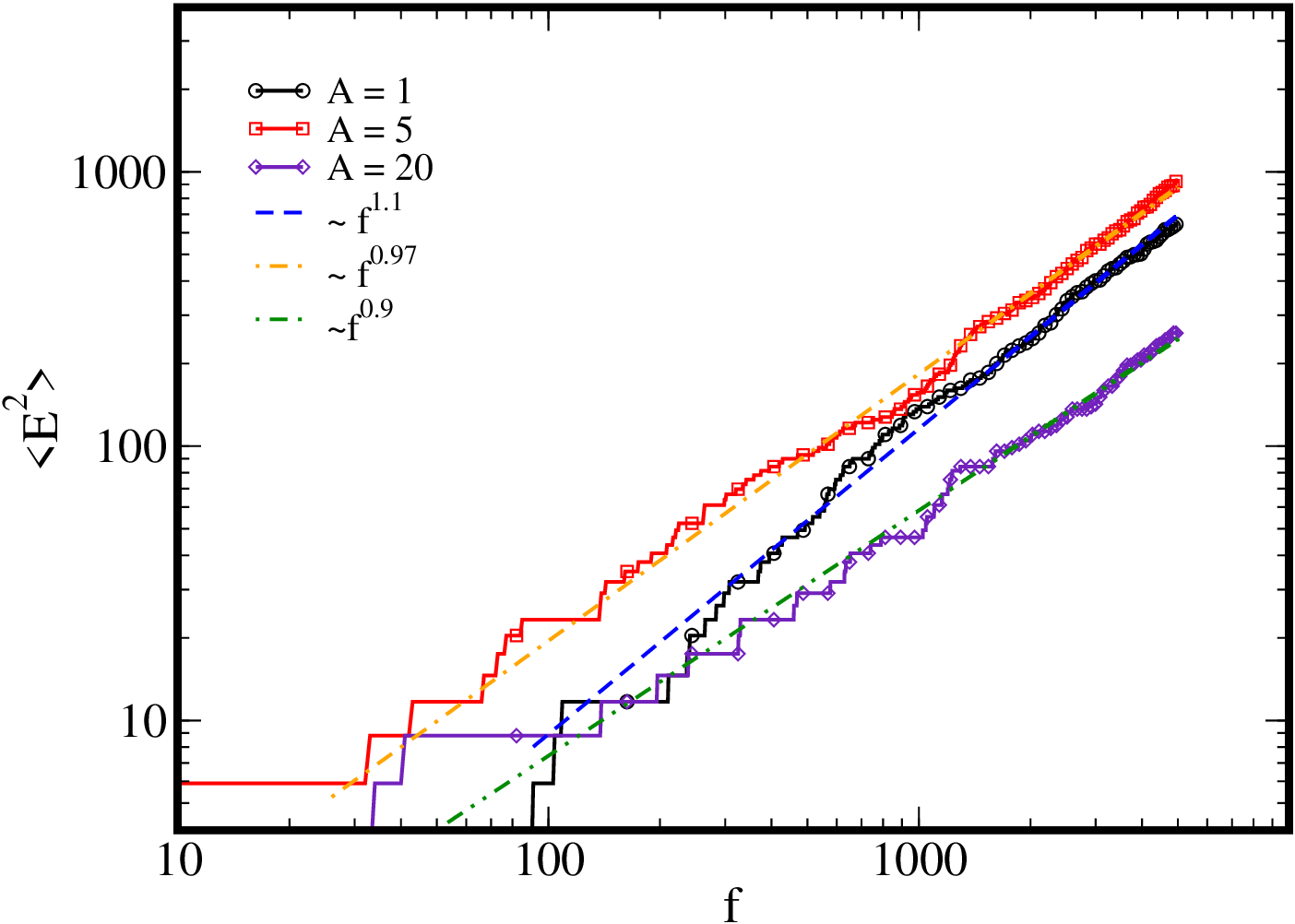}
\end{center}
\caption{The mean of the envy for a set of individuals in a closed system for two different distribution amplitudes $A = 1$ and $A = 5$ and non plastic preferences as a function of the comparison times $f$.}
\label{E1}
\end{figure}

\begin{figure}[th]
\begin{center}
\includegraphics[clip, width=0.6\textwidth]{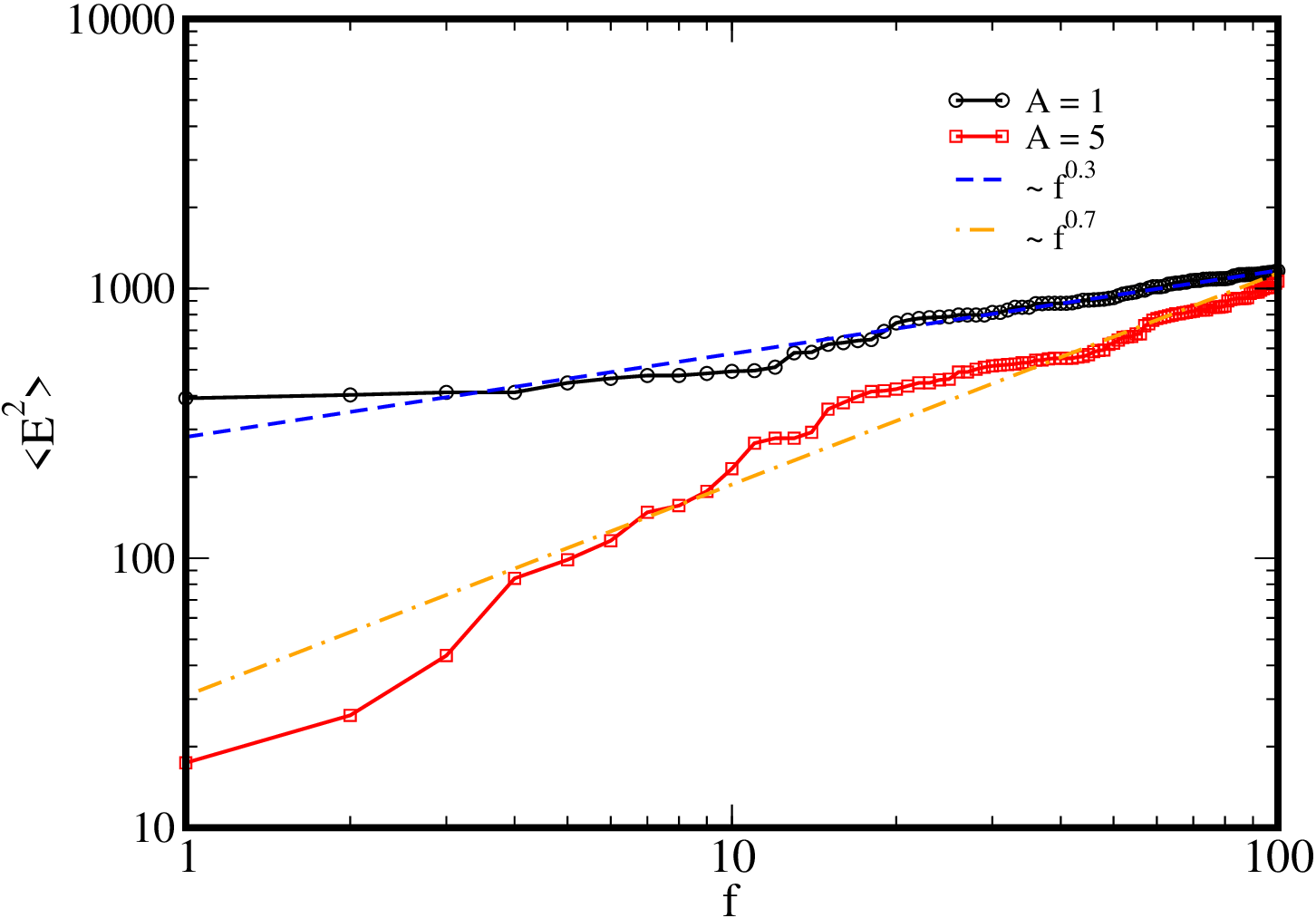}
\end{center}
\caption{Mean of the envy for a set of individuals in a closed system for two different distribution amplitudes $A = 1$ and $A = 5$ as a function of the comparisons times $f$. In this case  the individuals have ``plastic random'' preferences, i.e., each individual changes its preference list in a random way when she/he has not acquired the goods they want.}
\label{E2}
\end{figure}

The envy relation grows according to the increment of the number $f$ of interpersonal comparisons. A first analysis made for rigid preferences is shown in Fig. \ref{E1}. The fit of the curves show that $E \sim f^{1}$, which corresponds to a diffusive regime. That means, for non plastic preferences and infinite times the system relaxes in an statistical equilibrium with different levels of envy. However, when the preferences are plastic the system relaxes much faster (Fig. \ref{E2}). Here, the curves can be fitted with $E \sim f^{p}$, and $p < 1$. 

This results imply, the system is in a sub-diffusive regime, i.e, the system is subjected to dissipative forces (similar to the forces that produce an arrest of diffusing molecules in complex liquids \cite{Jensen}\cite{Diaz}), i.e., the introduction of random plastic preferences act as an additional force in the individuals, which avoids the free diffusion of envy among the system.

A very interesting scenario appears for the number of individuals expressing envy. With rigid preferences there is a relaxation process, from individuals expressing an initial anxiety to individuals that express some satisfaction level after an allocation process has been done (Fig. \ref{N2}). The autocorrelation function of density of individuals expressing envy can be fitted only for short times using an exponential function. However for intermediate time regimes the correlation function reaches a value equal to zero. Thereafter, this function relaxes to an equilibrium point. For this reason this curve can be fitted with an exponential function (typical for stochastic systems in equilibria) for  both short and upon intermediary and long time regimes. But the null values of the correlation function shows a tendency to a convolution of the correlation with a memory function (as is shown fitted in the insert of Fig. \ref{N2}), an effect that can be much stronger for individuals able to learn. Our interpretation is supported by the theory of correlation functions for complex liquids \cite{Yip}. We speculate that a system with non plastic preferences has a tendency to a non trivial relaxation function, i.e., individuals try to conserve their initial condition of non-envy. The distribution of goods is a force that drives this system, enforcing it to reach an statistical equilibrium.
  
\begin{figure}[th]
\begin{center}
\includegraphics[clip, width=0.6\textwidth]{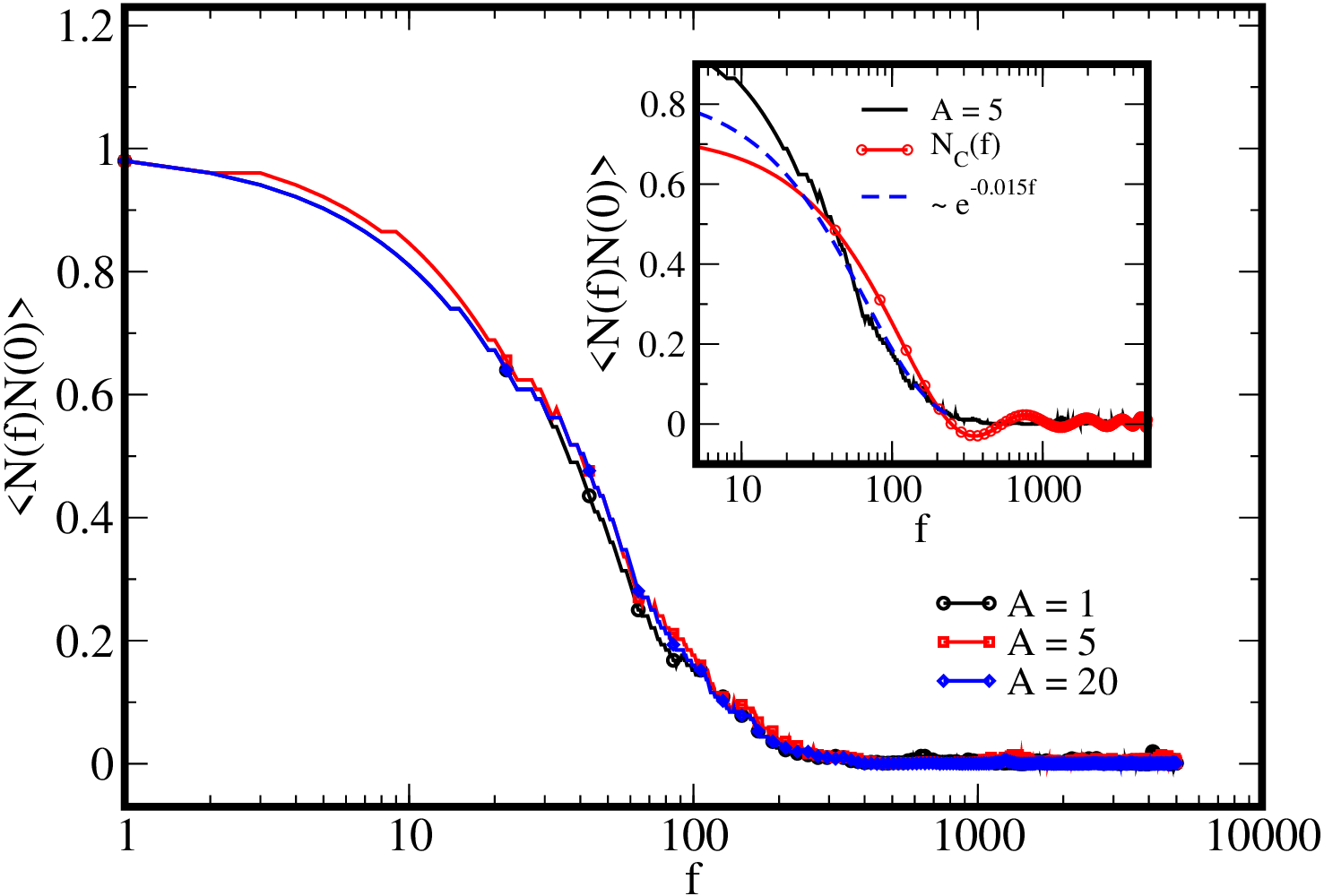}
\end{center}
\caption{Number of individuals feeling envy as a function of the comparisons time $f$. The first plot is a representation of the relaxation of individuals that express envy as a function of $f$. The insert shows the fit of this relaxation process for a correlation function with a negative contribution $N_{c}(f)$, typical for systems with memory functions.}
\label{N1}
\end{figure}

For plastic preferences this phenomenology is quite different. This situation is shown in Fig. \ref{N1}. Given that the individuals search for new preferences in each comparison time, the computational process is very intensive. For this reason, we performed simulations for short and intermediary time regimes. The must important feature in this results is that the relaxation process of the correlation function of envious individuals is suppressed and the system reaches in short comparisons time an equilibrium state. A comparison with the relaxation process of rigid preferences is also shown in the same plot. Therefore, we can see easily the positive effect of plastic preferences in the over relaxation process in the number of envious individuals weighted by comparisons time. It is a selective process that discards the initial non envious state and select the number of individuals that never reach a fairness state.

\section{Summary and Outlook}
The previous results are an alternative scenario to improve a system with individuals expressing envy. A very simple hypothesis considers that the problem of envy control only is related to an allocation mechanism. However, individuals posses also internal trends which disturb the way they express their own satisfaction level. Two cases has been considered in the present study: a first case where the individuals have rigid preferences and a second where the individuals can change their preferences if they do not become in a first instance what they want. In the worst of the cases, this adjust process can appears as a resignation of the individual to the circumstances. The effect of an elastic preference is the introduction of an over-relaxation process in the envy-dynamics of the system. The strong restriction of the present model is the consideration of a random process in the adjust of preferences. 

When elastic preferences are introduced, an uncertainty term is involved given by informational problems, implicit in social interaction systems, personal experiences and acquisition of new knowledge (i.e., reinforced or cognitive learning). To reduce its impact at the moment to choose, avoiding distortionary effects in their ultimate objective (welfare maximization), a cognitive learning mechanism must be introduced, allowing them to make better decisions in order to reach larger welfare levels regarding to their preferences. To remove the uncertainty component, a model testing for the ideal way and time to learning must be developed.

We thank Johannes J. Schneider for his support and enriching discussions preceding this paper and the physical support provided by the Complex Systems Lab of Prof. Bornholdt at University of Bremen.  


\end{document}